\documentstyle[epsf,12lomcon,cite]{article}

\def\babar{\mbox{\sl B\hspace{-0.4em} {\footnotesize\sl A}\hspace{-0.4em} B\hspace{-0.4em} {\footnotesize\sl A\hspace{-0.05em}R}}}
\def\eff{\varepsilon}

\def\ifb   {\mathrm{fb^{-1}}}

\def\epem  {\mathrm{e^+e^-}}

\def\GeV{\ifmmode{\mathrm{Ge\kern -0.1em V}}\else
                  \textrm{Ge\kern -0.1em V}\fi}
\def\MeV{\ifmmode{\mathrm{Me\kern -0.1em V}}\else
                  \textrm{Me\kern -0.1em V}\fi}

\newcommand{\eee}     {\ensuremath{e^-\!e^+\!e^-}}
\newcommand{\eemw}    {\ensuremath{\mu^+\!e^-\!e^-}}
\newcommand{\eemr}    {\ensuremath{\mu^-\!e^+\!e^-}}
\newcommand{\emmw}    {\ensuremath{e^+\!\mu^-\!\mu^-}}
\newcommand{\emmr}    {\ensuremath{e^-\!\mu^+\!\mu^-}}
\newcommand{\mmm}     {\ensuremath{\mu^-\!\mu^+\!\mu^-}}

\def\Nobs       {N_{\rm obs}}
\def\Nbgd       {N_{\rm bgd}}

\def\dEdM       {(\Delta {\rm{M}}, \Delta {\rm E})}
\def\mec        {\mathrm{m}_{\mathrm EC}}
\def\DeltaE     {\Delta {\rm{E}}}
\def\DeltaM     {\Delta {\rm{M}}}

\def\taumg      {\tau \to \mu \gamma}
\def\taueg      {\tau \to e \gamma}
\def\taulg      {\tau \to \ell \gamma}
\def\taulll     {\tau \to \ell \ell \ell}
\def\taulhh     {\tau \to \ell h h}

\def\BR         {{\cal{B}}}
\def\BRul       {{\cal{B}}_{\mathrm{UL}}^{90}}

\def\BRtaumg    {\BR(\taumg)}

\def\BRtaulg    {\BR(\taulg)}
\def\BRtaulll   {\BR(\taulll)}

\def\eg       {e^- \gamma}
\def\mg       {\mu^- \gamma}

\def\eee       {e^-   e^+  e^-}
\def\eemr      {\mu^- e^+  e^- }
\def\eemw      {\mu^+ e^-  e^- }
\def\emmr      {e^-  \mu^+ \mu^-}
\def\emmw      {e^+  \mu^- \mu^-}
\def\mmm       {\mu^- \mu^+ \mu^-}

\def\EKKr    {e^- K^+K^-}
\def\EKPr    {e^- K^+\pi^-}
\def\EPKr    {e^- \pi^+K^-}
\def\EPPr    {e^- \pi^+\pi^-}
\def\MKKr    {\mu^- K^+K^-}
\def\MKPr    {\mu^- K^+\pi^-}
\def\MPKr    {\mu^- \pi^+K^-}
\def\MPPr    {\mu^- \pi^+\pi^-}

\def\EKKw    {e^+ K^-K^-}
\def\EKPw    {e^+ K^-\pi^-}
\def\EPPw    {e^+ \pi^-\pi^-}
\def\MKKw    {\mu^+ K^-K^-}
\def\MKPw    {\mu^+ K^-\pi^-}
\def\MPPw    {\mu^+ \pi^-\pi^-}

\def\eett     {e^+e^- \to \tau^+\tau^-}
\def\eeqq     {e^+e^- \to q\bar{q}}

\begin{document}

\title{Lepton Flavor Violation in $\tau$ decays at \babar\
\footnote{To appear in the proceedings of
XII Lomonosov Conference on Elementary Particle Physics,
Moscow, Russia (25 - 31 August 2005).}}

\author{Swagato Banerjee \footnote{On behalf of the \babar\ Collaboration.}}

\address{University of Victoria, British Columbia, V8W 3P6 Canada.\\
E-mail: swaban@slac.stanford.edu}


\maketitle\abstracts{ 
Searches for lepton flavor violating 
$\taulg$, $\taulll$ and  $\taulhh$ decays
at the \babar\ experiment are presented.
Upper limits on the branching ratios are obtained
at the level of $O(10^{-7})$ at 90\% confidence level.}

\section{Introduction}
\label{sec:intro}

Lepton flavor violating (LFV) processes such as the neutrinoless decay of the $\tau$ lepton
have long been identified as unambiguous signatures of new physics,
because no known fundamental local gauge symmetry forbids such a decay.
While forbidden in the Standard Model (SM) because of vanishing neutrino mass in the three lepton generations, 
extensions to include current knowledge of neutrino mass and mixing imply $\BRtaumg \sim {\cal{O}} (10^{-54})$~\cite{Aubert:2005ye},
which is many orders of magnitude below the experimental sensitivity.
However, many new theories, as tabulated below,
allow for LFV decays: $\taulg$, $\taulll$, $\taulhh$ (where $\ell = e, \mu; h = \pi, K$)
up to their existing experimental bounds $\sim {\cal{O}} (10^{-7})$:\\
\vspace*{-.5cm}
\begin{table}[!h]
\begin{center}
\begin{tabular}{l|c|c}
                                                    & {$\BRtaulg$} & {$\BRtaulll$}\\\hline
mSUGRA + seesaw~\cite{Ellis:1999uq,Ellis:2002fe}    & $10^{-7}$    & $10^{-9}$  \\
SUSY + SO(10)~\cite{Masiero:2002jn,Fukuyama:2003hn} & $10^{-8}$    & $10^{-10}$ \\
SM + seesaw~\cite{Cvetic:2002jy}                    & $10^{-9}$    & $10^{-10}$ \\
Non-Universal Z$^\prime$~\cite{Yue:2002ja}          & $10^{-9}$    & $10^{-8}$  \\
SUSY + Higgs~\cite{Dedes:2002rh,Brignole:2003iv}    & $10^{-10}$   & $10^{-7}$  \\
\end{tabular}
\end{center}
\end{table}
\vspace*{-.25cm}

Feynman diagrams for $\tau \to \mu \gamma$ and $\tau \to \mu \mu \mu$ decays 
via s-neutrino mixing in minimal supergravity model with heavy $\nu_{\rm{R}}$ (seesaw mechanism) 
and via neutral Higgs exchange in supersymmetric model
are shown in Figure 1, respectively.

\begin{figure}[!h]
\begin{center}
\begin{minipage}[l]{.4\textwidth}
\begin{center}
\epsfxsize=\textwidth\epsfysize=.15\textheight\epsfbox{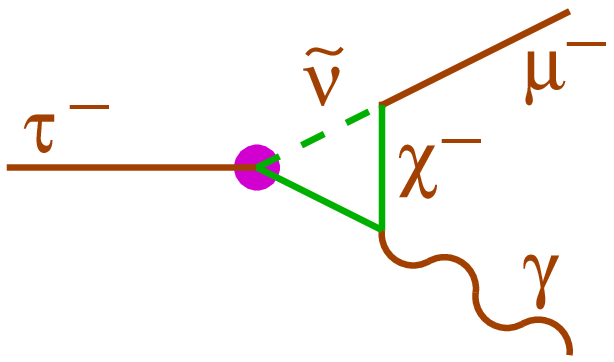}
\end{center}
\end{minipage}
\hspace*{.1\textwidth}
\begin{minipage}[r]{.4\textwidth}
\begin{center}
\epsfxsize=\textwidth\epsfysize=.15\textheight\epsfbox{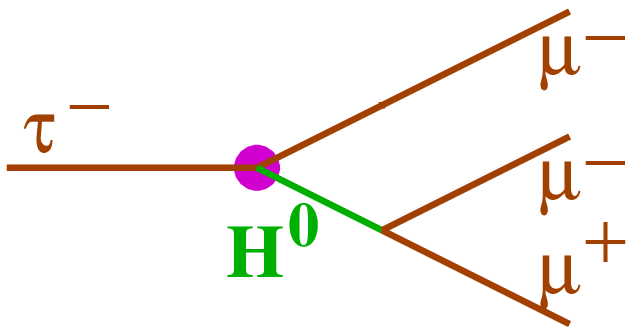}
\end{center}
\end{minipage}
\end{center}
Figure 1: Illustrative scenarios for $\tau \to \mu \gamma$ (left) and $\tau \to \mu \mu \mu$ (right).
\end{figure}

\section{Signal Identification}
\label{sec:signal}

Searches for LFV decay in modes: $\taulg$~\cite{Aubert:2005ye,Aubert:2005wa}, $\taulll$~\cite{Aubert:2003pc} 
and $\taulhh$~\cite{Aubert:2005tp} have been performed with 232.2 $\ifb$, 91.6 $\ifb$ and 221.4 $\ifb$
of data collected by the \babar\ experiment at $\sqrt{s}$ $\approx$ 10.58 \GeV, respectively.
The characteristic feature of these decays is that both the energy and 
the mass of the $\tau$-daughters are known in such an $\epem$ annihilation environment.
In terms of the two independent variables:
beam energy constrained mass $(\mec)$ and the energy variable ${\DeltaE} = \mathrm{E}_{\tau} - \sqrt{s}/2$,
where $\mathrm{E}_{\tau}$ is the energy of the $\tau$-daughters in center-of-mass system,
the signal is clustered around $(m_\tau,0)$ in the ($\mec,\DeltaE$) plane.

The identification of daughters from signal $\tau$ decays are optimized
for searches for each decay mode separately. The electrons are identified
from the energy deposited in the electromagnetic calorimeter and 
momentum of the track measured in the drift chamber 
with an efficiency of 91\% for $\taueg$ and $\taulll$ searches. 
The muons are identified by its minimal ionizing particle signature in the calorimeter
and hits in the instrumented flux return with an efficiency of 82\%, 63\% and 44\% in 
$\taumg$, $\taulll$ and $\taulhh$ searches respectively. The kaons are identified 
using the measured rate of ionization loss in the drift chamber and 
the measured Cherenkov angle in a ring-imaging detector with an efficiency of 81\%.
The mis-identification rates for a pion track to be identified as an electron, a muon or a kaon
are 0.1\%, 1.0 $-$ 4.8\%, 1.4\% respectively.

\section{Background estimation}
\label{sec:bkg}

The primary backgrounds are from Bhabha or di-muon events, 
which are restricted to a narrow band at small values of $|\DeltaE|$,
or the $\eett$ events, which are restricted to negative values of $\DeltaE$,
because the signal topology reconstruction does not account for the missing neutrino's.
The remaining backgrounds from $\eeqq$ are uniformly distributed.

\begin{figure}[!h]
\begin{center}
\begin{minipage}[l]{.49\textwidth}
\begin{center}
\epsfxsize=\textwidth\epsfysize=.240\textheight\epsfbox{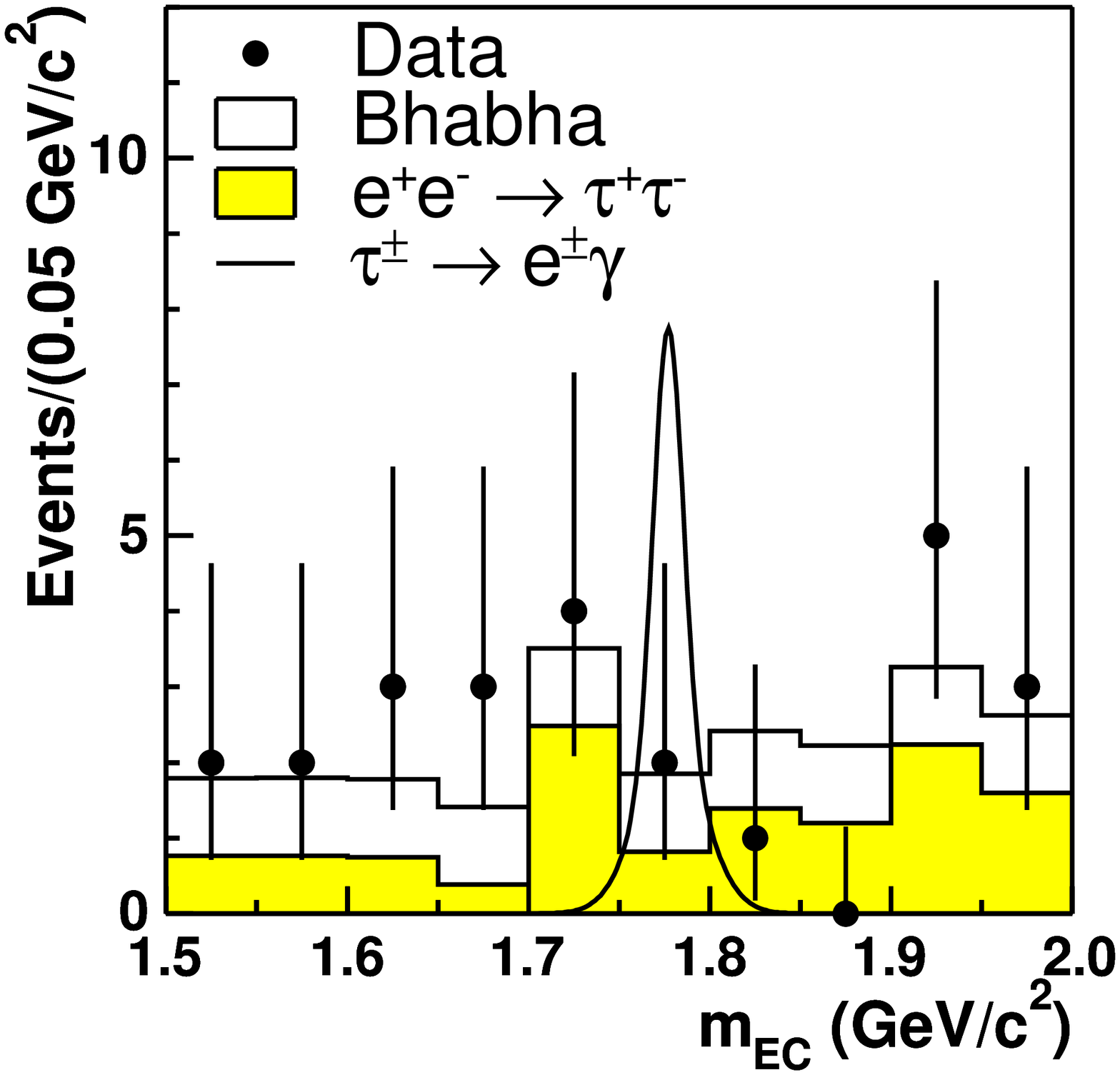}
\end{center}
\end{minipage}
\begin{minipage}[r]{.49\textwidth}
\begin{center}
\epsfxsize=\textwidth\epsfysize=.230\textheight\epsfbox{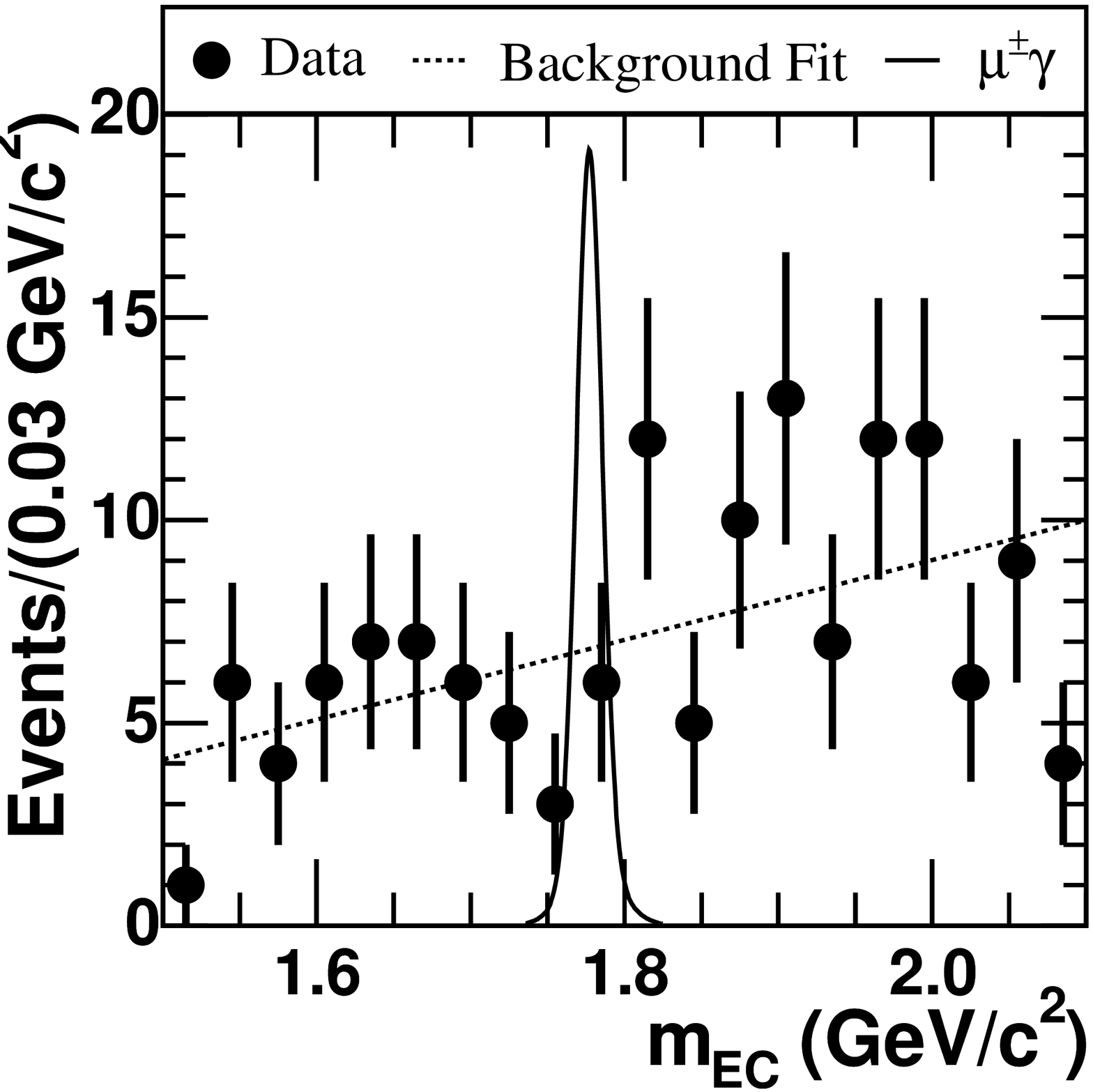}
\end{center}
\end{minipage}
\end{center}
Figure 2: $\mec$ distribution inside a $\pm 2 \sigma$ band in $\DeltaE$
for $\taulg$ searches.
\end{figure}

For $\taulg$ searches, the signal probability density function (PDF) is described 
by a double Gaussian shape in $\mec$, and the background is well described by a constant PDF
or with a small slope in $\mec$ inside a $\pm 2 \sigma$ band in $\DeltaE$,
as shown in Figure 2 for $\taueg$ (left) and $\taumg$ (right) decays.

For $\tau \to \ell \ell \ell ~(\ell h h)$ searches, the background PDF's are analytically parameterized
as function of $\DeltaE$ and $\DeltaM$ = m$_{dau}$ - m$_\tau$, where
m$_{dau}$ is the reconstructed mass of the $\tau$ daughters.
The background rates are determined by un-binned maximum likelihood fits to the data,
shown along with the selected signal MC events in Figure 3 and 4
for $\taulll$ and $\taulhh$ decay modes respectively.

All these searches are performed in a blinded manner, where the background predictions from sideband data are compared 
to the data inside the signal region, only after the optimization and systematic studies have been completed.

\begin{figure}[!h]
\begin{center}
\begin{minipage}[c]{.75\textwidth}
\begin{center}
\epsfxsize=\textwidth\epsfysize=.24\textheight\epsfbox{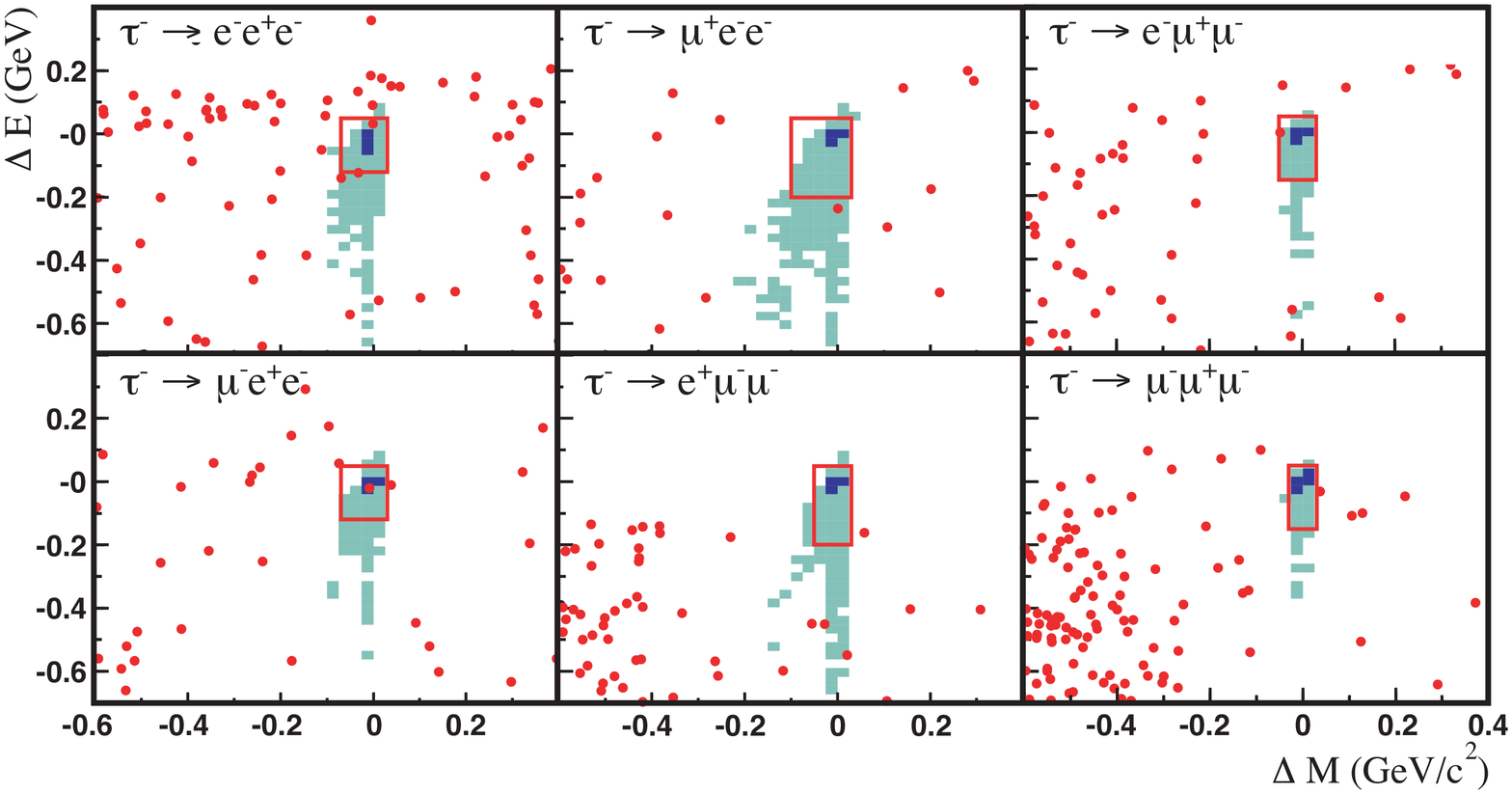}
\end{center}
\end{minipage}
\end{center}
Figure 3: Observed data as dots and the boundaries of the signal region in $\dEdM$ plane for $\taulll$ searches.
The dark and light shading indicates contours containing 50\% and 90\% of the selected MC signal events, respectively.
\end{figure}

\vspace*{-.75cm}
\begin{figure}[!h]
\begin{center}
\begin{minipage}[c]{\textwidth}
\begin{center}
\epsfxsize=\textwidth\epsfysize=.24\textheight\epsfbox{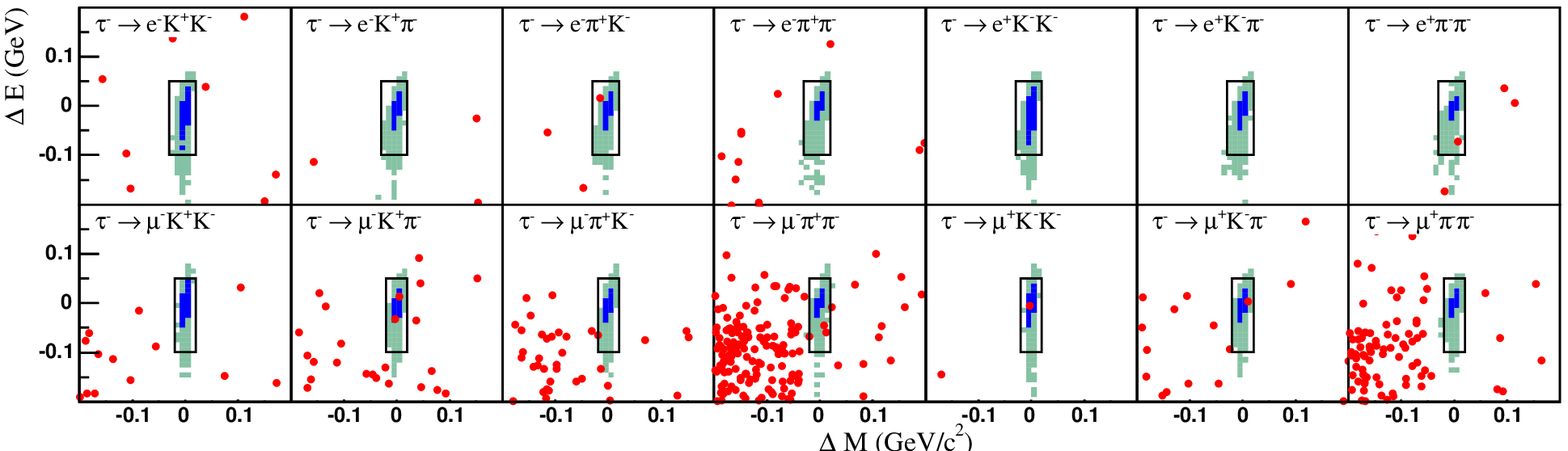}
\end{center}
\end{minipage}
\end{center}
Figure 4: Observed data as dots and the boundaries of the signal region in $\dEdM$ plane for $\taulhh$ searches.
The dark and light shading indicates contours containing 50\% and 90\% of the selected MC signal events, respectively.
\end{figure}

\section{Results}
\label{sec:res}

No signal has been observed.
Upper limits at 90\% confidence level (C.L.) are set using:
$\BR^{90}_{UL}=N^{90}_{UL}/(2\eff {\cal{L}} \sigma_{\tau\tau})$,
where $N^{90}_{UL}$ is the 90\% C.L. upper limit on the number of signal events for $\Nobs$ events observed
when $\Nbgd$ background events are expected, and $\eff$ is the signal efficiency.
Efficiency estimates, the number of expected background events $(\Nbgd)$ in the signal region (with total uncertainties),
the number of observed events $(\Nobs)$ in the signal region, and the 90\% C.L. upper limit $(\BRul)$ for each decay mode
are tabulated below:\\
\vspace*{-.5cm}
\begin{table}[!h]
\begin{center}
\begin{tabular}{lcccc}\hline\hline
Mode    & Efficiency [\%]& $\Nbgd$         & $\Nobs$ & $\BRul (10^{-7})$ \\\hline
$\eg$	&$ 4.7 \pm 0.3 $ & $ 1.9 \pm 0.4 $ & 1       & 1.1 \\
$\mg$	&$ 7.4 \pm 0.7 $ & $ 6.2 \pm 0.5 $ & 4       & 0.7 \\\hline
$\eee$	&$ 7.3 \pm 0.2 $ & $ 1.5 \pm 0.1 $ & 1       & 2.0 \\
$\eemw$	&$11.6 \pm 0.4 $ & $ 0.4 \pm 0.1 $ & 0       & 1.1 \\
$\eemr$	&$ 7.7 \pm 0.3 $ & $ 0.6 \pm 0.1 $ & 1       & 2.7 \\
$\emmw$	&$ 9.8 \pm 0.5 $ & $ 0.2 \pm 0.1 $ & 0       & 1.3 \\
$\emmr$	&$ 6.8 \pm 0.4 $ & $ 0.4 \pm 0.1 $ & 1       & 3.3 \\
$\mmm$	&$ 6.7 \pm 0.5 $ & $ 0.3 \pm 0.1 $ & 0       & 1.9 \\\hline
$\EKKr$	&$ 3.8 \pm 0.2 $ & $ 0.2 \pm 0.1 $ & 0       & 1.4 \\
$\EKPr$	&$ 3.1 \pm 0.1 $ & $ 0.3 \pm 0.1 $ & 0       & 1.7 \\
$\EPKr$	&$ 3.1 \pm 0.1 $ & $ 0.1 \pm 0.1 $ & 1       & 3.2 \\
$\EPPr$	&$ 3.3 \pm 0.2 $ & $ 0.8 \pm 0.1 $ & 0       & 1.2 \\
$\MKKr$	&$ 2.2 \pm 0.1 $ & $ 0.2 \pm 0.1 $ & 0       & 2.5 \\
$\MKPr$	&$ 3.0 \pm 0.2 $ & $ 1.7 \pm 0.3 $ & 2       & 3.2 \\
$\MPKr$	&$ 2.9 \pm 0.2 $ & $ 1.0 \pm 0.2 $ & 1       & 2.6 \\
$\MPPr$	&$ 3.4 \pm 0.2 $ & $ 3.0 \pm 0.4 $ & 3       & 2.9 \\
$\EKKw$	&$ 3.9 \pm 0.2 $ & $ 0.0 \pm 0.0 $ & 0       & 1.5 \\
$\EKPw$	&$ 3.2 \pm 0.1 $ & $ 0.2 \pm 0.1 $ & 0       & 1.8 \\
$\EPPw$	&$ 3.4 \pm 0.2 $ & $ 0.4 \pm 0.1 $ & 1       & 2.7 \\
$\MKKw$	&$ 2.1 \pm 0.1 $ & $ 0.1 \pm 0.1 $ & 1       & 4.8 \\
$\MKPw$	&$ 2.9 \pm 0.2 $ & $ 1.5 \pm 0.3 $ & 1       & 2.2 \\
$\MPPw$	&$ 3.3 \pm 0.2 $ & $ 1.5 \pm 0.3 $ & 0       & 0.7 \\\hline\hline
\end{tabular}
\end{center}
\end{table}

\section{Summary}
\label{sec:sum}
An improvement of five order of magnitude in the upper limits for 4 LFV $\tau$ decay modes
over the past two decades is shown in Figure 5.
The next five years promises to be most interesting phase in this evolution,
when experimental results approach closer the predictions from different theoretical models.

\begin{figure}[!h]
\begin{center}
\begin{minipage}[l]{.495\textwidth}
\begin{center}
\epsfxsize=\textwidth\epsfysize=.22\textheight\epsfbox{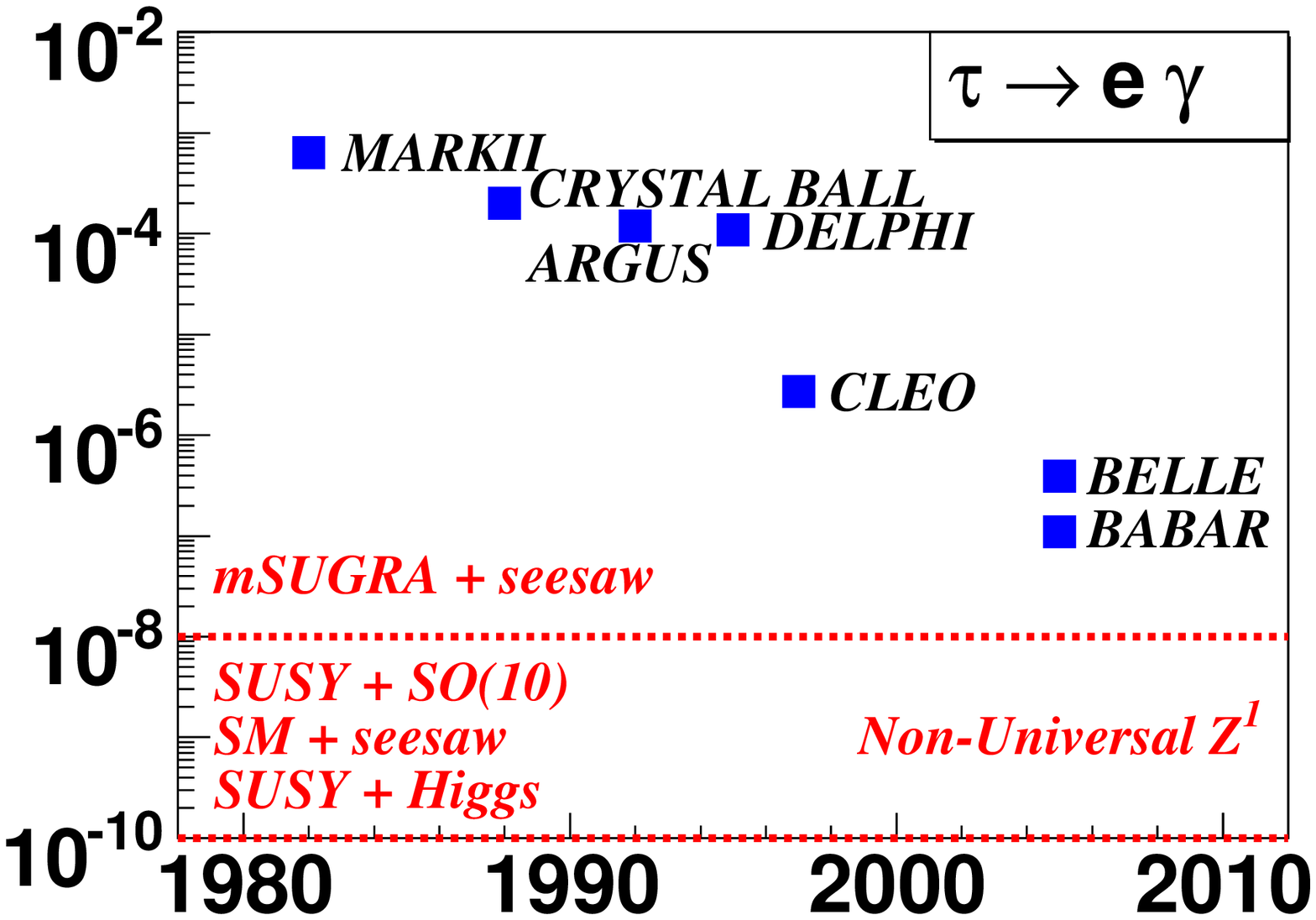}
\end{center}
\vspace*{-.5cm}
\end{minipage}
\begin{minipage}[r]{.495\textwidth}
\begin{center}
\epsfxsize=\textwidth\epsfysize=.22\textheight\epsfbox{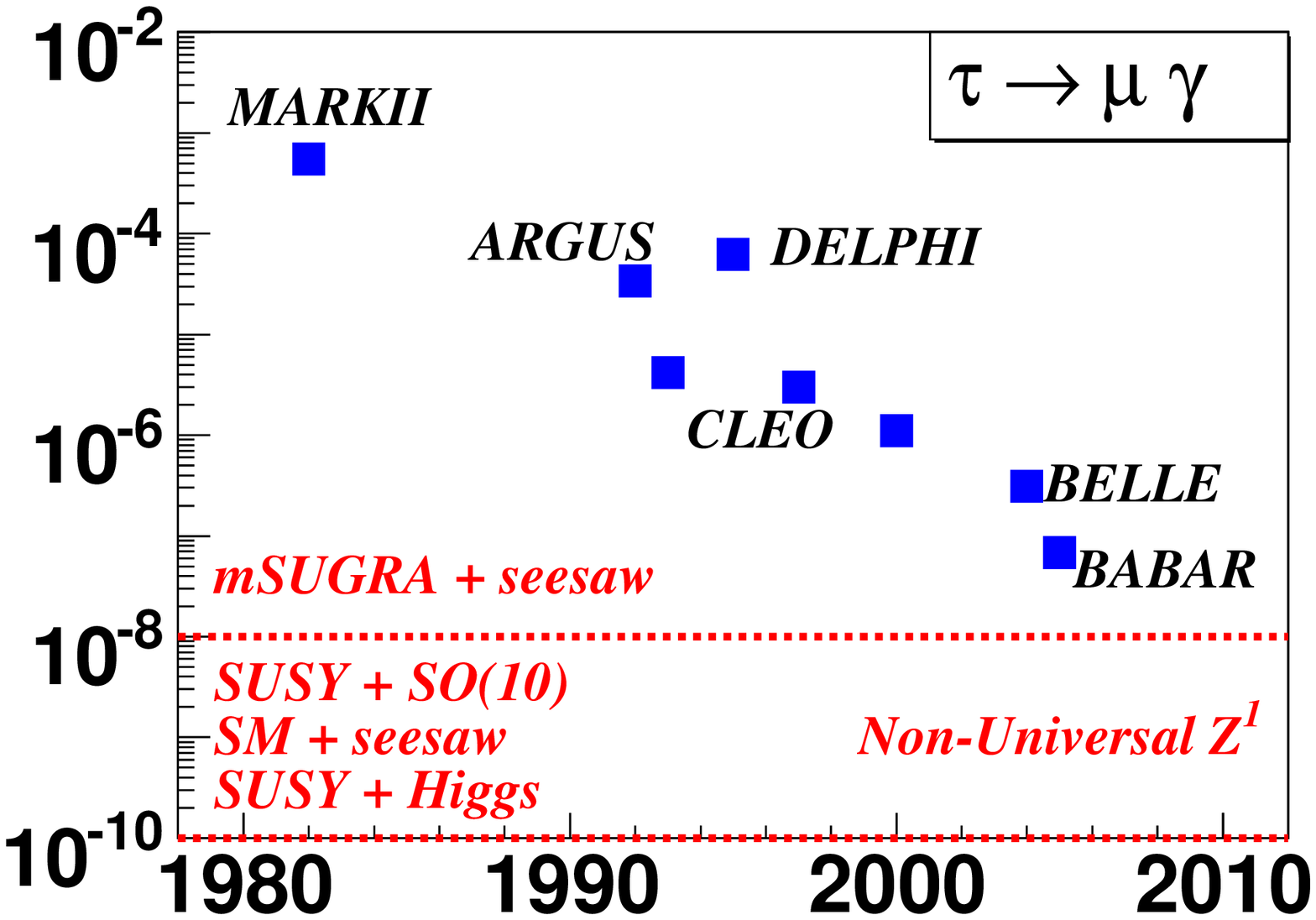}
\end{center}
\vspace*{-.5cm}
\end{minipage}
\begin{minipage}[l]{.495\textwidth}
\begin{center}
\epsfxsize=\textwidth\epsfysize=.22\textheight\epsfbox{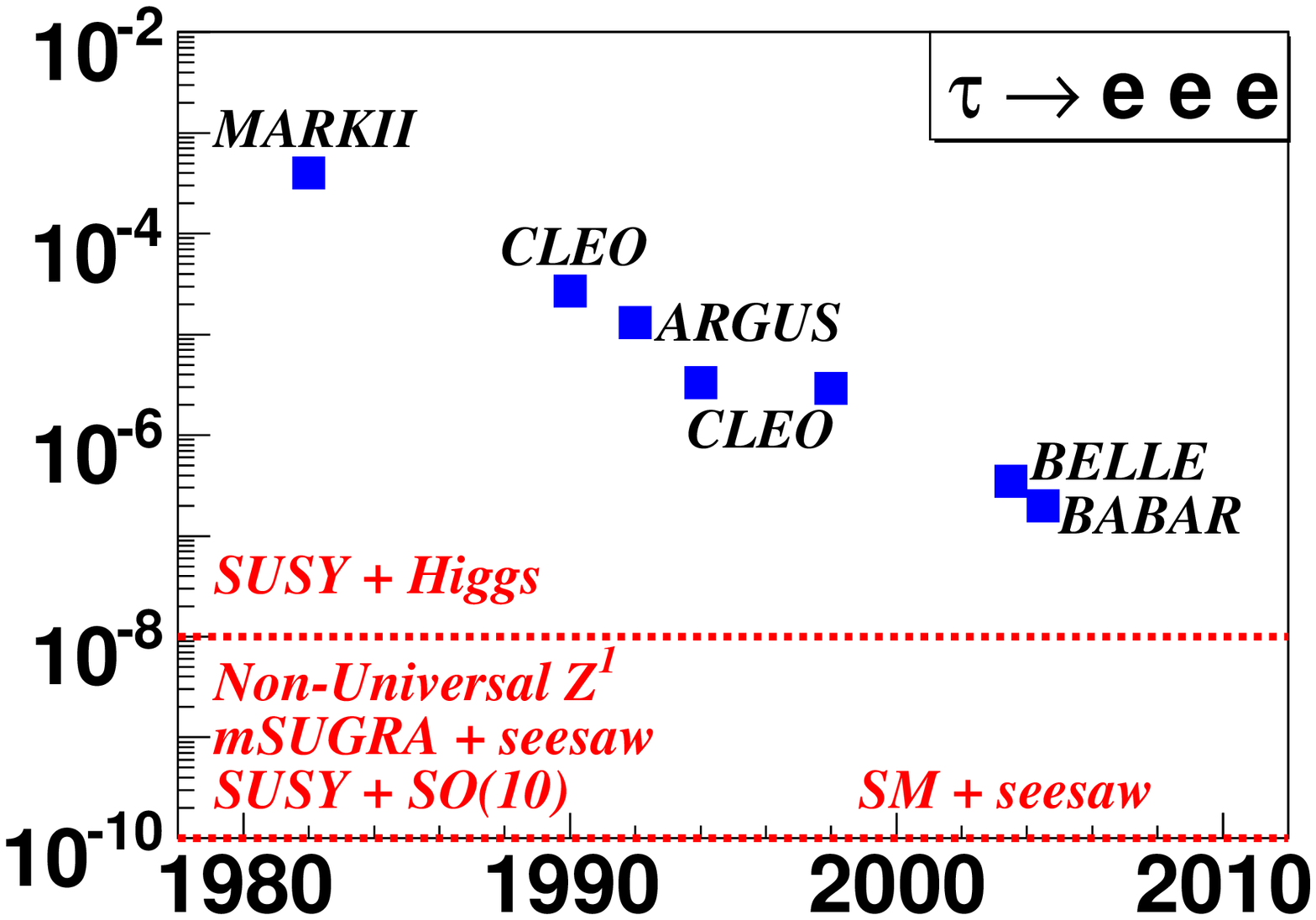}
\end{center}
\vspace*{-.5cm}
\end{minipage}
\begin{minipage}[r]{.495\textwidth}
\begin{center}
\epsfxsize=\textwidth\epsfysize=.22\textheight\epsfbox{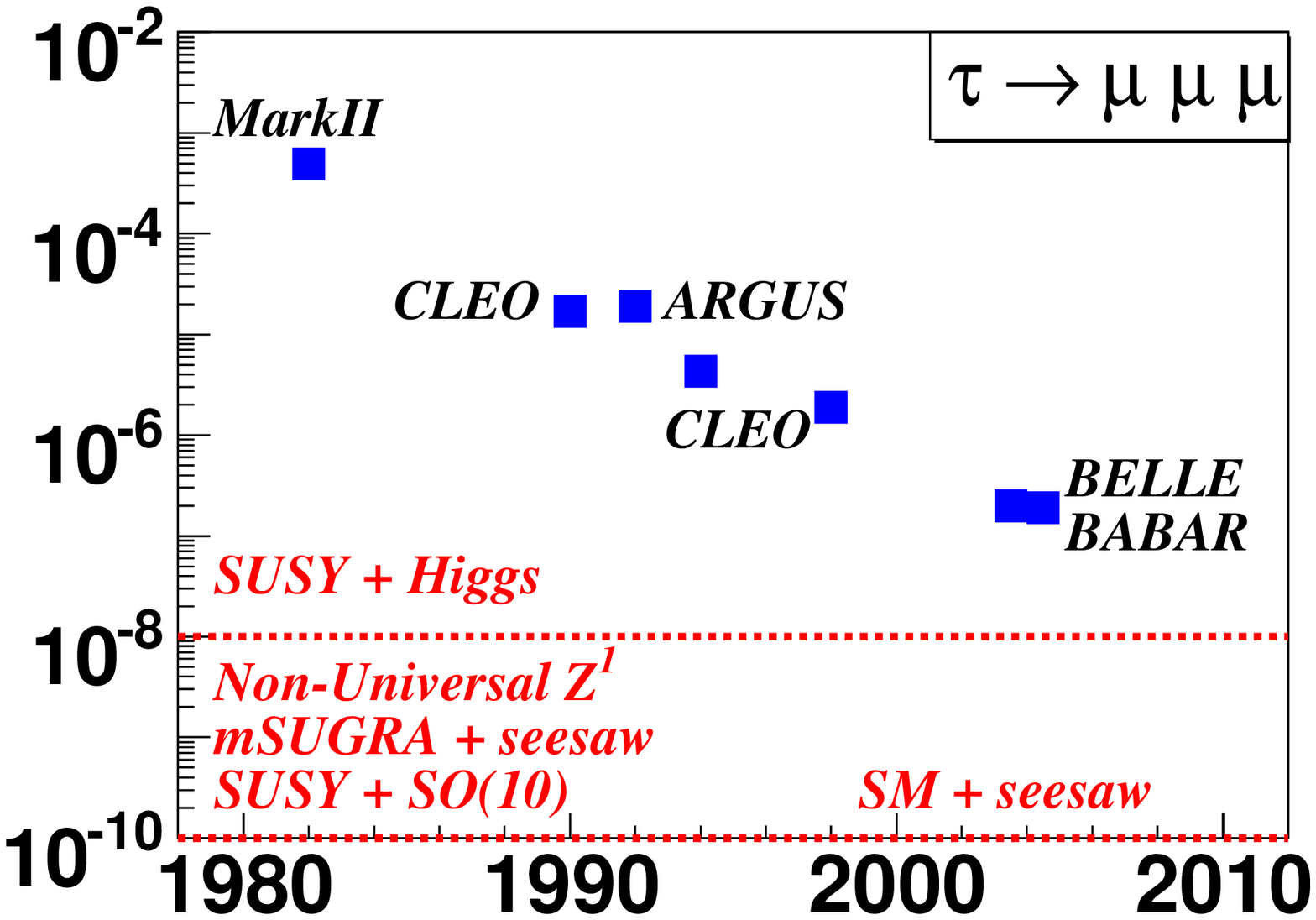}
\end{center}
\vspace*{-.5cm}
\end{minipage}
\end{center}
Figure 5: Evolution of experimental bounds ($\BRul$) and some predictions.
\end{figure}


\section*{References}
\begin{thebibliography}{99}
\vspace*{-.25cm}
\bibitem{Aubert:2005ye}
  B.~Aubert {\it et al.}  [\babar\ Collaboration],
  Phys.\ Rev.\ Lett.\  {\bf 95}, 041802 (2005).


\bibitem{Ellis:1999uq}
  J.~R.~Ellis, M.~E.~Gomez, G.~K.~Leontaris, S.~Lola, D.~V.~Nanopoulos,
  Eur.\ Phys.\ J.\ C {\bf 14}, 319 (2000).

\bibitem{Ellis:2002fe}
  J.~R.~Ellis, J.~Hisano, M.~Raidal, Y.~Shimizu,
  Phys.\ Rev.\ D {\bf 66}, 115013 (2002).

\bibitem{Masiero:2002jn}
  A.~Masiero, S.~K.~Vempati, O.~Vives,
  Nucl.\ Phys.\ B {\bf 649}, 189 (2003).

\bibitem{Fukuyama:2003hn}
  T.~Fukuyama, T.~Kikuchi, N.~Okada,
  Phys.\ Rev.\ D {\bf 68}, 033012 (2003).

\bibitem{Cvetic:2002jy}
  G.~Cvetic, C.~Dib, C.~S.~Kim, J.~D.~Kim,
  Phys.\ Rev.\ D {\bf 66}, 034008 (2002)
  [Erratum-ibid.\ D {\bf 68}, 059901 (2003)].

\bibitem{Yue:2002ja}
  C.~x.~Yue, Y.~m.~Zhang, L.~j.~Liu,
  Phys.\ Lett.\ B {\bf 547}, 252 (2002).

\bibitem{Dedes:2002rh}
  A.~Dedes, J.~R.~Ellis, M.~Raidal,
  Phys.\ Lett.\ B {\bf 549}, 159 (2002).

\bibitem{Brignole:2003iv}
  A.~Brignole, A.~Rossi,
  Phys.\ Lett.\ B {\bf 566}, 217 (2003).


\bibitem{Aubert:2005wa}
  B.~Aubert {\it et al.}  [\babar\ Collaboration],
  arXiv:hep-ex/0508012. Submitted to  Phys.\ Rev.\ Lett.\  (2005).

\bibitem{Aubert:2003pc}
  B.~Aubert {\it et al.}  [\babar\ Collaboration],
  Phys.\ Rev.\ Lett.\  {\bf 92}, 121801 (2004).

\bibitem{Aubert:2005tp}
  B.~Aubert {\it et al.}  [\babar\ Collaboration],
  Phys.\ Rev.\ Lett.\  {\bf 95}, 191801 (2005).




\end{thebibliography}
\end{document}
